\documentclass[aps,prc,reprint,superscriptaddress,floatfix,showpacs,nofootinbib]{revtex4-1}

\usepackage{epsfig}
\usepackage{dcolumn}
\usepackage{bm}
\usepackage{amssymb}
\usepackage{amsmath}

\usepackage{mathtools}
\usepackage{mathrsfs}

\usepackage{amsfonts}
\usepackage{soul}
\usepackage[usenames,dvipsnames]{color}

\newcommand{\tvec}[1]{\vec{#1}_\bot}

\begin{document}

\title{Initializing BSQ with Open-Source ICCING}

\author{Patrick Carzon}
\email[Email: ]{pcarzon2@illinois.edu}
\affiliation{Illinois Center for Advanced Studies of the Universe, Department of Physics, University of Illinois at Urbana-Champaign, Urbana, IL 61801, USA}
\author{Mauricio Martinez}
\email[Email: ]{mmarti11@ncsu.edu}
\affiliation{North Carolina State University, Raleigh, NC 27695, USA}
\author{Matthew D. Sievert}
\email[Email: ]{msievert@nmsu.edu}
\affiliation{Department of Physics, New Mexico State University, Las Cruces, NM 88003, USA}
\author{Douglas E. Wertepny}
\affiliation{(formerly) Ben-Gurion University of the Negev, Beer-Sheva 84105, Israel}
\author{Jacquelyn Noronha-Hostler}
\email[Email: ]{jnorhos@illinois.edu}
\affiliation{Illinois Center for Advanced Studies of the Universe, Department of Physics, University of Illinois at Urbana-Champaign, Urbana, IL 61801, USA}

\date{\today}
\begin{abstract}
While it is well known that there is a significant amount of conserved charges in the initial state of nuclear collisions, the production of these due to gluon splitting has yet to be thoroughly investigated. The ICCING (Initial Conserved Charges in Nuclear Geometry) algorithm reconstructs these quark distributions, providing conserved strange, baryon, and electric charges, by sampling a given model for the $g \rightarrow q\bar{q}$ splitting function over the initial energy density, which is valid at top collider energies, even when $\mu_B=0$. The ICCING algorithm includes fluctuations in the gluon longitudinal momenta, a structure that supports the implementation of dynamical processes, and the c++ version is now open-source. A full analysis of parameter choices on the model has been done to quantify the effect these have on the underlying physics. We find there is a sustained difference across the different charges that indicates sensitivity to hot spot geometry.
\end{abstract}

\maketitle

%
\section{Introduction}
%

In principle, initial conditions consist of a complete specification of the full initial energy-momentum tensor and initial conserved charge currents for baryon number $B$, strangeness $S$, and electric charge $Q$. In practice, only the initial energy density is used and all other components are set to zero, though there has been recent progress on including more initial state variables \cite{Gardim:2011qn, Gardim:2012yp, Gale:2012rq, Schenke:2019pmk, Liu:2015nwa, Kurkela:2018wud,Plumberg:2021bme,Chiu:2021muk}. Some thought has been given to conserved charge currents, primarily finite net baryon density \cite{Werner:1993uh, Itakura:2003jp, Shen:2017bsr, Akamatsu:2018olk, Mohs:2019iee}, which is motivated by the search for a critical point in the Quantum Chromodynamics (QCD) phase diagram in this regime \cite{Karpenko:2015xea, Borsanyi:2018grb, Noronha-Hostler:2019ayj, Monnai:2019hkn, Monnai:2016kud, Bazavov:2018mes, Critelli:2017oub, Parotto:2018pwx, Demir:2008tr, Denicol:2013nua, Denicol:2018wdp, Kadam:2014cua, Stephanov:1999zu, Stephanov:2017ghc, Jiang:2015hri, Mukherjee:2016kyu, Nahrgang:2018afz, An:2019osr, Du:2019obx, Batyuk:2017sku, Shen:2017bsr}. At top collider energies, a mean field assumption is made that the total charge in the plasma vanishes which motivates setting the charge density $\rho$ uniformly to zero. This assumption is consistent with a leading-order picture in perturbative QCD which describes the initial state as composed of only gluons. Event-by-event fluctuations in the energy-momentum tensor have been established as essential to initial conditions of heavy-ion collisions \cite{Takahashi:2009na,Alver:2010gr}, this mean field assumption ignores local fluctuations of charge density around zero.

At next-to-leading order in pQCD, the gluons are able to pair produce quarks and antiquarks in equal proportions which allows for local fluctuations in charge density while preserving a total charge of zero. This has been studied in the proton where $q \bar{q}$ pairs, as sea quarks, are a significant contribution to the parton distribution functions (PDFs) as compared to gluons. This is illustrated by the gluon, $x g$, and sea quark, $x S$, PDFs extracted by the H1 and ZEUS Collaborations in \cite{Aaron:2009aa}. The parameter $x$ is proportional to the inverse of the collision energy, and so low values correspond to the energies present in heavy-ion collisions. In the small $x$ regime, the gluon and sea quark distributions scale proportionately, $x S \propto x g$, and this is consistent with the perturbative expectation, $x S \approx \alpha_s \, x g$. Sea quark fluctuations can come from other nonperturbative mechanisms \cite{Shuryak:2002qz} as well.

In order to study the transport of charge at top collider energies, an initial condition generator that contains $q \bar{q}$ pairs and a hydrodynamic code that can propagate all three conserved charges are needed. This work focuses on the implementation of the former as a Monte Carlo algorithm that takes any initial condition, that can be described as an energy density, and samples the $g \rightarrow q \bar{q}$ splitting probability to introduce BSQ conserved charges as quark pairs. The algorithm presented here is ICCING, Initial Conserved Charges in Nuclear Geometry, and is formulated in a modular way to allow for the selection of different choices for all theoretically dependent components, including the initial energy density input, the $g \rightarrow q \bar{q}$ splitting probabilities, and the distribution of charge densities \cite{Carzon:2019qja,Martinez:2019jbu}. The events modified by ICCING can be read directly into hydrodynamic simulations. To illustrate the algorithm and its physical consequences, the Trento initial condition generator \cite{Moreland:2014oya, Bass:2017zyn, Moreland:2018gsh}, is chosen for its ability to match experimental results and its model agnostic construction, and the $g \rightarrow q \bar{q}$ spatial correlation functions derived in Ref.~\cite{Martinez:2018ygo} are used.

%
\section{Model}

%
%
\begin{figure}
\begin{center}
	\includegraphics[width=0.45\textwidth]{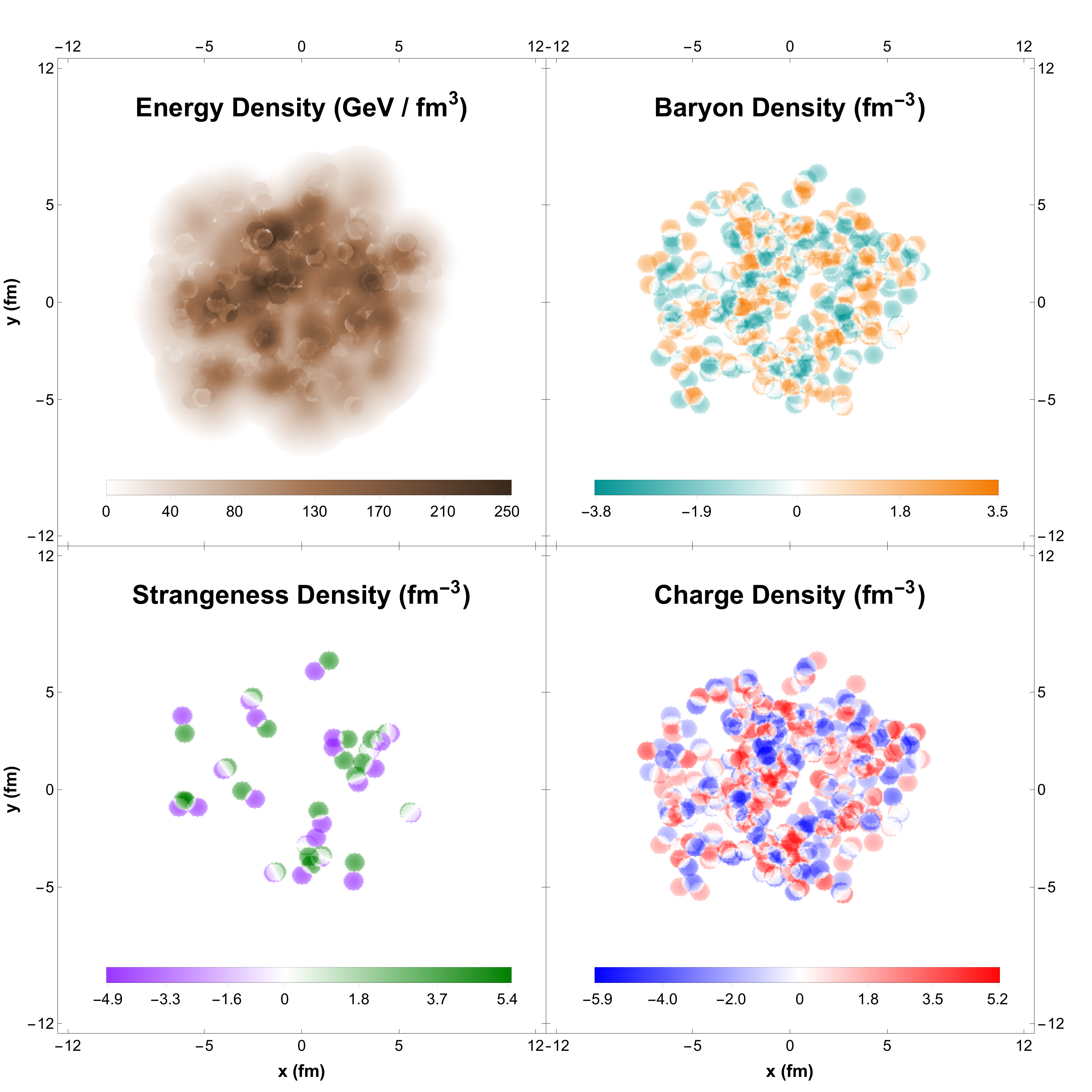}
    \caption{An event after being sampled by the ICCING algorithm, resulting in a reconstructed energy density as well as new distributions of the three conserved charges.  Note that some artifacts of the energy redistribution can be seen in the modified energy density. Taken from \cite{Carzon:2019qja}.
	}
	\label{f:Events}
	\end{center}
\end{figure}
%

Here I present a high level overview of the ICCING algorithm. On an event basis, the first step is an externally provided energy density profile $\epsilon (\vec{x}_\bot)$ and some profile that may be used as the saturation scale. From this initial profile, a valued point is chosen randomly to act as the center of a possible gluon and the energy available, inside a circle of a predefined radius, is calculated. A distribution is sampled to determine the fraction of the available energy that will be the gluon. Then a Monte Carlo sampling of the quark flavor ratios provided in \cite{Carzon:2019qja} is used to determine whether the gluon will split into a quark antiquark pair and what its flavor would be if this were the case. Another sampling is done to determine both the distance between the two quarks and the fraction of energy that is shared by each (the probability distributions for this process are beyond the scope of this work and are detailed in Ref.~\cite{Carzon:2019qja}). Finally, the energy, baryon, strange, and charge densities of the quark and antiquark are distributed in the output density grids and if the gluon did not undergo a splitting, its energy is copied over without modification. This process is repeated until all of the energy from the input density has been transferred to the output grids. A plot of the densities from a fully processed event is shown in Fig.~\ref{f:Events}.

The probability of gluons to split into quark/antiquark pairs and the reconstruction of the initial condition, including the conserved charge densities, is specified through external input allowing flexibility in theoretical dependence. This allows ICCING to be agnostic as to the choice of the quark/gluon splitting function. We make a particular choice, for the analysis here, to use the quark/gluon multiplicities obtained from a color-glass condensate (CGC) calculation for this input. The theoretical calculation underlying this choice was presented in Ref.~\cite{Martinez:2018ygo}.

An expression for the differential probability distribution $\frac{dP}{dr_\bot \, d\alpha}$ for a gluon to split into a $q \bar q$ pair with given kinematics is the result of the calculation in Ref.~\cite{Martinez:2018ygo}.  Here $\tvec{r}$ is the separation vector between the quark and antiquark, and $\alpha$ is the fraction of light-front momentum (i.e., energy) carried by the quark. Integrating over this differential probability with respect to $r_\bot$ and $\alpha$ gives the total probability for a gluon to split into quarks of a given flavor.  The calculation of Ref.~\cite{Martinez:2018ygo} gives an expression for the probability distribution in terms of dipole scattering amplitudes, which are natural degrees of freedom in CGC effective theory. We have evaluated the resulting probability distribution for the Golec-Biernat-Wusthoff (GBW) \cite{Golec-Biernat:1998zce} model of the dipole amplitude.

%
\section{Results}
%

    %
    \begin{figure} 
    \begin{centering}
    	\includegraphics[width=\linewidth]{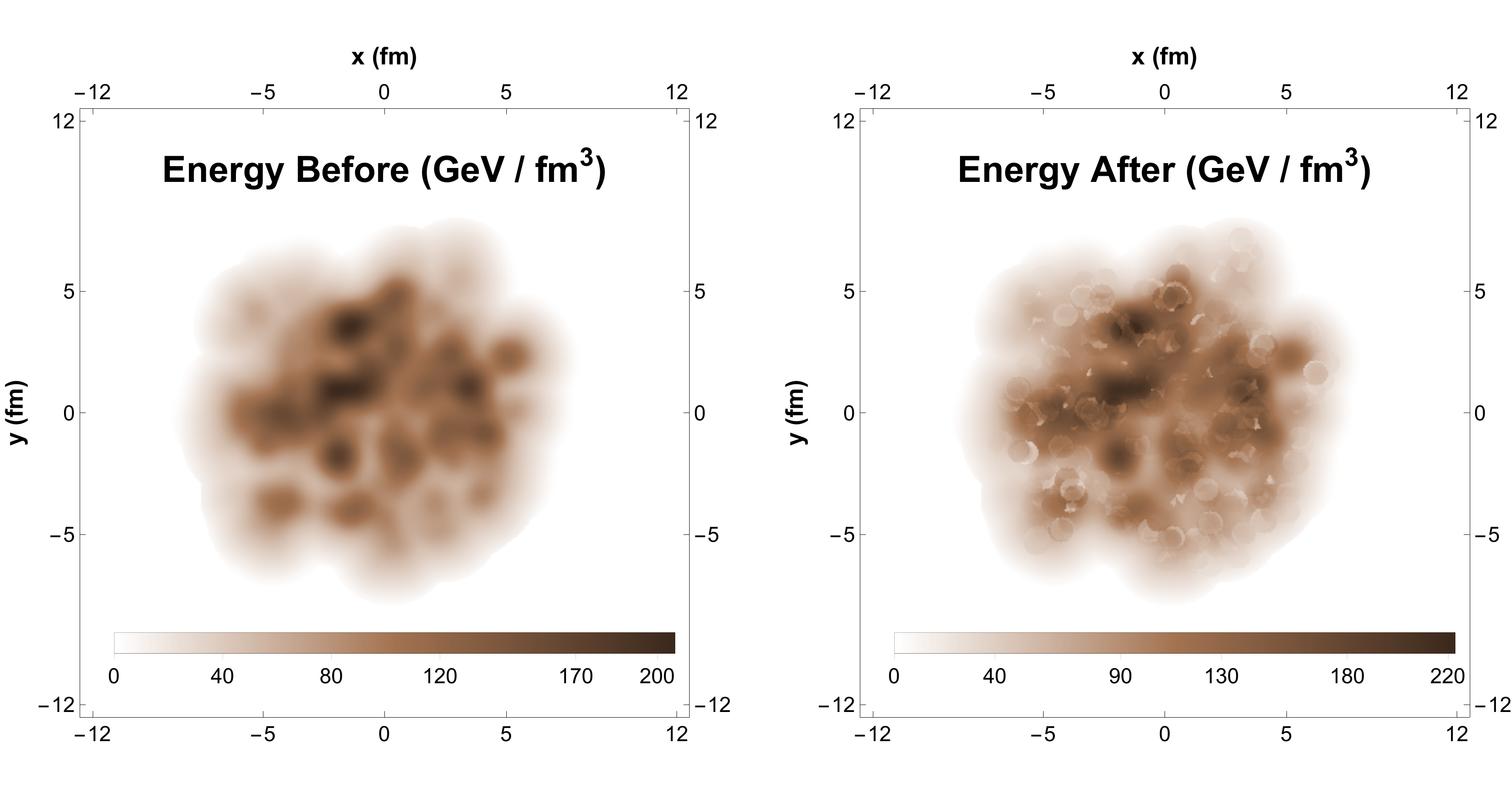}
    \end{centering}
    	\caption{
    	Comparison of the energy density before (left plot) and after (right plot) running the ICCING algorithm.  As a result of redistributing the energy density from the $g \rightarrow q \bar{q}$ splitting, the energy density profile is somewhat modified, including visible artifacts associated with the model implementation. Taken from \cite{Carzon:2019qja}.}
    	\label{f:BeforeAfterPlot}

    \end{figure}
    
    %

It is important to quantify the extent to which the ICCING algorithm modifies the input energy density through the redistribution of energy due to the quantum mechanical $g\to q\bar{q}$ splitting process. There is nontrivial modification of the energy density by this process as see in Fig.~\ref{f:BeforeAfterPlot}, where it is possible to see circular artifacts in the energy density where gluons were sampled and split into quarks. Though the artifacts are highly dependent on artificial choices made by the selection of parameter values, i.e. the gluon radius $r$, they are a real effect of the partonic splitting physics. The sampling artifacts are important to quantify since Trento, by itself, has already been shown to accurately describe final state flow harmonics and fluctuations of bulk particle production at the LHC \cite{Moreland:2014oya, Bernhard:2016tnd, Alba:2017hhe, Giacalone:2017dud, Bernhard:2019bmu}. Consistency with experimental data could be disrupted if ICCING introduces new charge distributions at the cost of modifying the energy profile too much.

To check the modification of the energy density profile, we estimate the effect on the flow harmonics. 
The final state is often described by harmonic flow coefficients, $v_n$, which are constructed from correlations between particles produced by the collision. These flow harmonics reflect the response of the collision system to the underlying physics present. The final state flow coefficients have been shown \cite{Niemi:2012aj} to be related to the initial state eccentricities nearly linearly:
\begin{equation}
v_n=\kappa_n \varepsilon_n .
\end{equation}
This relation is explored in more detail in the work I did in Ref.~\cite{Carzon:2020xwp} where we explored deformations of Pb to attempt explaining the $v_2$ to $v_3$ puzzle. This relationship is not perfect and in peripheral collisions non-linear corrections to the hydrodynamic response can become significant \cite{Noronha-Hostler:2015dbi, Sievert:2019zjr, Rao:2019vgy}.

    %
     \begin{figure}
        \begin{centering}
    	\includegraphics[width=0.45\textwidth]{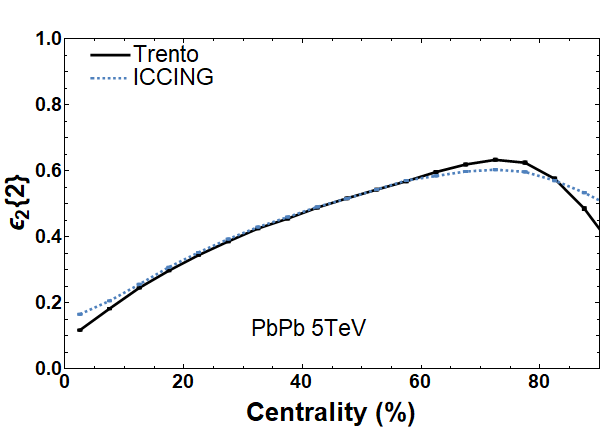} \,
    	\includegraphics[width=0.45\textwidth]{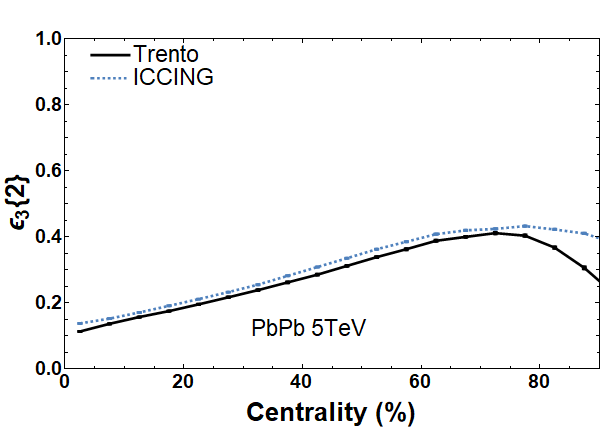}
    	\end{centering}
    	\caption{Eccentricities of the energy density as a function of centrality before (solid curve) and after running ICCING (dotted curve). Taken from \cite{Carzon:2019qja}.}
    	\label{f:EccInOut}
    \end{figure}
   %
The 2-particle cumulants of ellipticity and triangularity of the energy density as a function of centrality, shown in Fig.~\ref{f:EccInOut}, are a good way to quantify the effect of ICCING on the original initial condition. Modification of the event geometry by ICCING is very small for $\varepsilon_2 \{2\}$ in the $10-60\%$ centrality range and there is a consistently constant increase for $\varepsilon_3 \{2\}$ in the same window. This shows that the effect of gluon splitting, as introduced, is smaller than that of the average elliptical geometry of the collision, but does introduce a new source of small event-by-event fluctuations. There is an increase in ellipticity for the most central events, $0-10\%$ centrality. The nucleon-level geometry in these collisions tends to be quite round, and it appears that ICCING converts these geometries into something more elliptical due to the $g \rightarrow q \bar{q}$ splitting naturally producing quarks that are back-to-back. This is also seen in peripheral collisions $\sim 70-90\%$, where the peak of $\varepsilon_2 \{2\}$ shifts to the right. Previous work \cite{Sievert:2019zjr} has shown that the location of this peak characterizes a transition between impact-parameter-driven geometry and finite-number-driven geometry characterized by $N_{part}$ at the level of Trento. Thus, the location of the peak reflects a resolution scale for the constituents of the medium, and a modification of this transition due to the introduction of sub-nucleonic degrees of freedom in ICCING is natural. It is important to note that the final-state anisotropic flow, $v_n$, in more peripheral collisions will be modified by effects such as nonlinear response to the initial geometry \cite{Noronha-Hostler:2015dbi,Sievert:2019zjr}. It is reasonable to conclude that modifications to the bulk energy geometry introduced by ICCING are minimal, with interesting systematic differences in very central and very peripheral events.

%
\subsection{Strangeness As a Distinct Probe of the Initial State}
%
    %
    \begin{figure}
        \begin{centering}
    	\includegraphics[width=0.45 \textwidth]{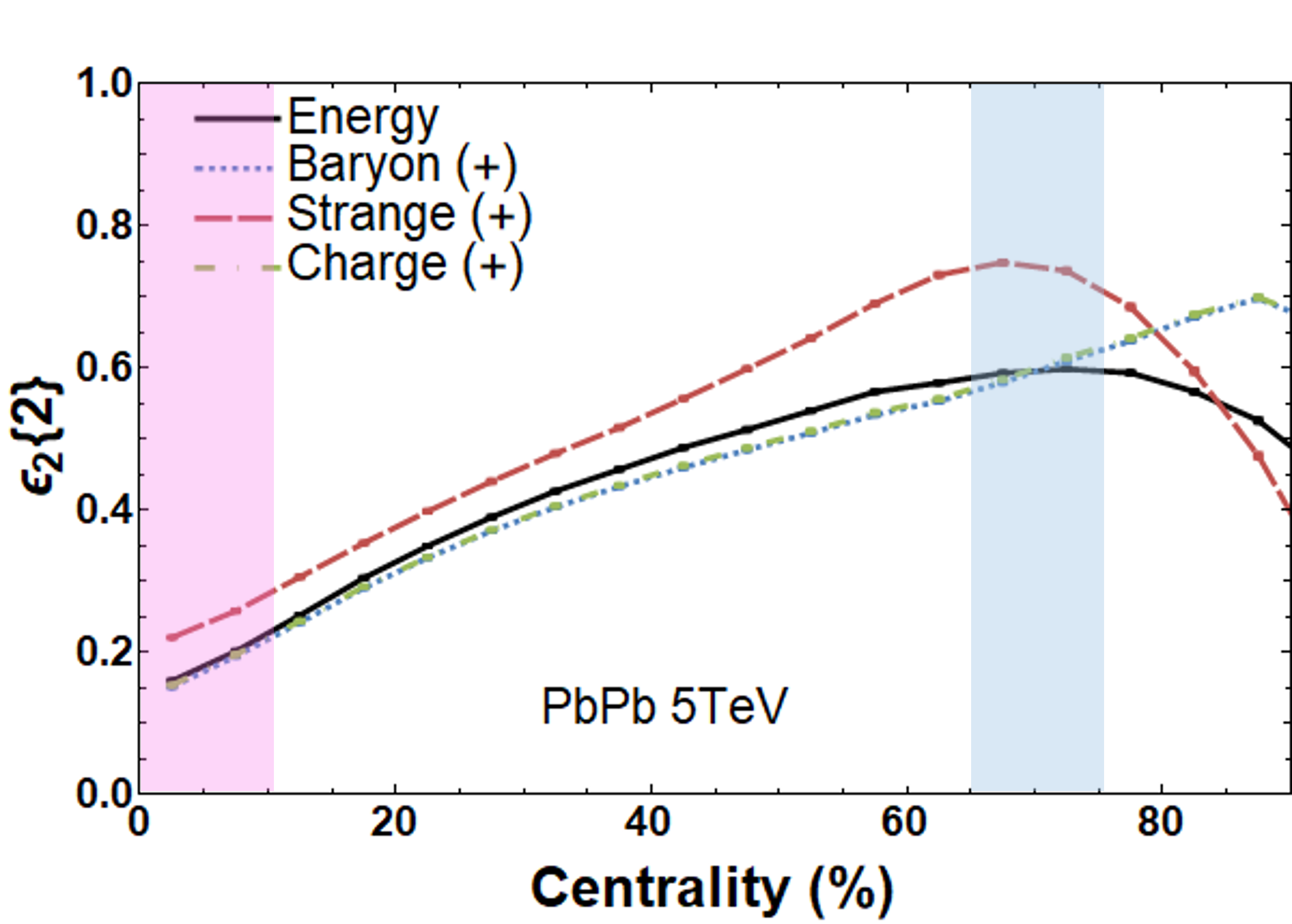}
    	\end{centering}
    	\caption{
    	RMS ellipticity $\varepsilon_2 \{2\}$ versus centrality.  Two centrality bins $(0-10\%)$ and $(65-75\%)$ are highlighted with the corresponding histograms shown below. Taken from \cite{Carzon:2019qja}.
    	}
    	\label{f:IBSQCentrality}
    \end{figure}
    %

Having established the effect of ICCING on the initial condition input, we can look at the geometries of charge produced by the algorithm. We run the initial conditions using Trento PbPb collisions at $\sqrt{s_{NN}}=5.02$ TeV.  One must choice an normalization constant and equation of state to obtain the initial energy density from Trento (that is given as an entropy density). We use the PDG16+/2+1[WB] equation of state and normalization from \cite{Alba:2017hhe}.

 \begin{figure}
        \begin{centering}
    	\vspace{.5cm}
    	\includegraphics[width=0.45 \textwidth]{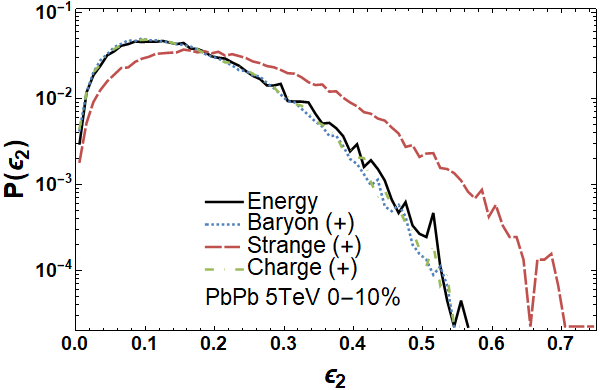} \,
    	\includegraphics[width=0.45 \textwidth]{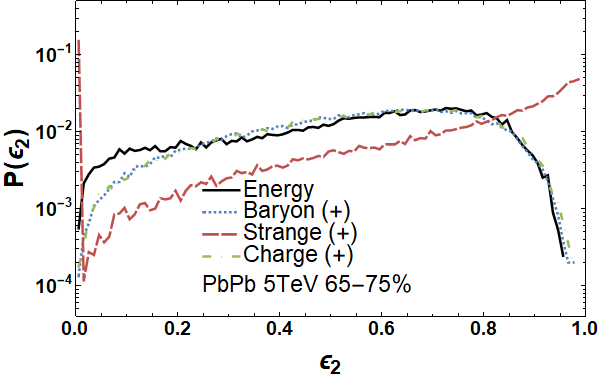}
    	\end{centering}
    	\caption{
    	Top: Distribution of ellipticity in the $0-10\%$ bin.
    	Bottom: Distribution of ellipticity in the $65-75\%$ bin. Taken from \cite{Carzon:2019qja}.
    	}
    	\label{f:IBSQdistro}
    \end{figure}

The elliptic eccentricities of energy, baryon, strangeness, and electric charge are shown in Fig.~\ref{f:IBSQCentrality}. In  Fig.~\ref{f:IBSQCentrality} is $\varepsilon_2 \{2\}$, where, for $ < 65\%$ centrality, the baryon number and electric charge distributions track the energy distribution very closely, while the strangeness distribution is much more eccentric. To gain a better intuition of this behaviour, we examine the central $0-10\%$ centrality bin (pink band) differentially by plotting the corresponding probability distribution in the top plot of Fig.~\ref{f:IBSQdistro}. On an event-by-event basis, the $B, Q$ distributions produce the same geometry profile as the energy density; we can attribute this to the huge abundances of $u, d$ quarks produced in these central events. These nearly-massless quarks are produced in large quantities and are nearly homogeneous in central collisions, so that the resulting charge densities mirror the original energy density profile, quantitatively.  In contrast, the strangeness distribution is more eccentric than the bulk geometry, even in the most central collisions.  This indicates that the strange quarks do not saturate the collision, instead generating a lumpy and highly anisotropic geometry even when the energy is quite round.

Another area that would be important to analyze differentially are peripheral collisions at $65-75\%$ centrality, where we see the $B, Q$ ellipticities begin to deviate from the bulk energy density and the strangeness experiences its peak in $\varepsilon\left\{2\right\}$. This centrality range is highlighted by the blue band and is shown in the bottom panel of Fig.~\ref{f:IBSQdistro}. Looking at the $B, Q$ distributions, we see that they qualitatively track the same geometry as the energy density, reflecting the still-dominant role of impact-parameter-driven geometry, but nontrivial differences start to appear. Events start to shift to larger $B, Q$ eccentricities when compared to the energy distributions that produced them, indicating the $u, d$ quark abundances have dropped low enough to no longer saturate the bulk geometry and are now characterizing a different, lumpier distribution. The strangeness distribution for this centrality window has a dramatic increase in events with strangeness ellipticity close to $1$ and a large spike about zero. These peaks in the distribution, at $0$ and $1$, most likely reflect a events with few strange quarks, where exactly one or two $s \bar{s}$ pairs may be created, which leads to one or two blobs of positive strangeness that are innately round (zero eccentricity) or innately elliptical (maximal eccentricity). This double-peak behavior in $\epsilon_2^{(S+)}$ correlates with the location of the peak in the cumulant $\varepsilon_2^{(S+)} \{2\}$, again indicative that there is a transition to event geometries controlled by a small number of produced strange quarks.

    %
    \begin{figure}
        \begin{centering}
    	\includegraphics[width=0.45 \textwidth]{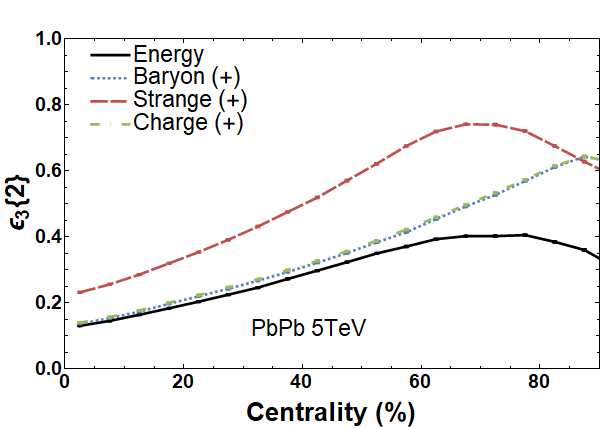}
    	\end{centering}
    	\caption{
        The triangular eccentricity, $\varepsilon_3 \{2\}$, versus centrality. Taken from \cite{Carzon:2019qja}.
    	}
    	\label{f:E3_BSQCentrality}
    \end{figure}
    %

The triangularity,$\varepsilon_3 \{2\}$, of energy and $BSQ$ densities is shown in Fig.~\ref{f:E3_BSQCentrality}. Unlike $\varepsilon_2 \{2\}$ which is dominated by the mean-field elliptical geometry below $\lesssim 60\%$ centrality, $\varepsilon_3 \{2\}$ arises entirely from fluctuations, without this mean-field background.  Up to very peripheral collisions $> 80\%$ centrality, there is a clear hierarchy in the densities: $\varepsilon_3^{(S^+)} \{2\}  > \varepsilon_3^{(B^+ \, , \, Q^+)} \{2\} > \varepsilon_3^{(E)} \{2\}$. This is consistent with the introduction of new sources of sub-nucleonic fluctuations contributing to the $BSQ$ distributions, since strange quarks are produced the least and therefore their geometry fluctuates the most. Triangularity, $\varepsilon_3$, is the most sensitive probe into the differences between the various charge and energy geometries due to its direct sensitivity to the fluctuating charge distributions, independent of the mean-field background.

    %
    \begin{figure}[h!]
    	\includegraphics[width=0.45 \textwidth]{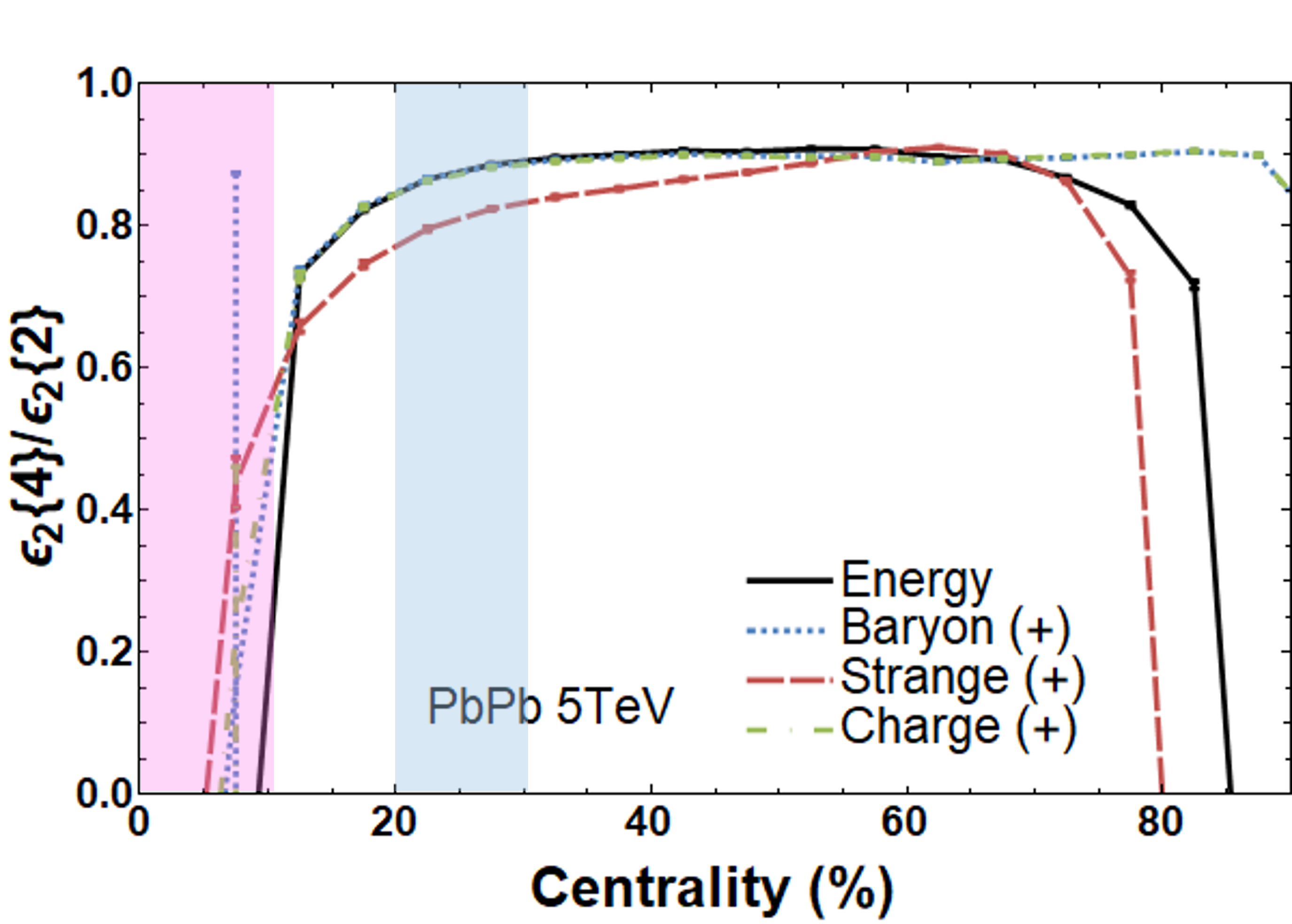}
    	\\
    	\vspace{.5cm}
    	\includegraphics[width=0.45 \textwidth]{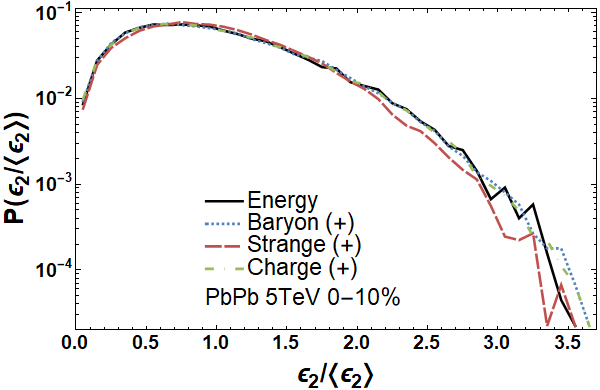}
    	\includegraphics[width=0.45 \textwidth]{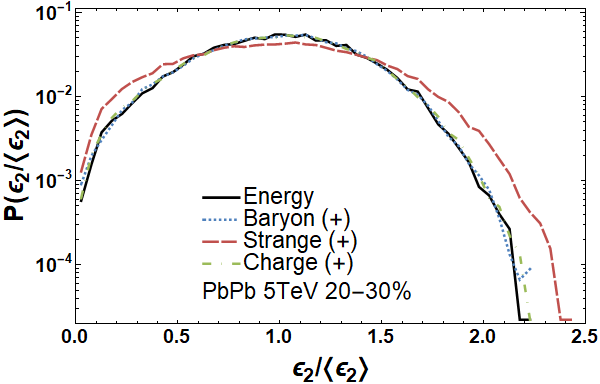}
    	\caption{
    	\textbf{Top:} Cumulant ratio $\varepsilon_2 \{4\} / \varepsilon_2 \{2\}$ versus centrality.  Two centrality bins ($0-10\%$ and $20-30\%$) are highlighted, corresponding to the histograms below.
    	\textbf{Bottom Left:} Probability distribution of the ellipticity $\varepsilon_2$ in the $0-10\%$ bin, scaled by the mean $\langle \varepsilon_2 \rangle$ to illustrate the width of the distribution.  Note that the strangeness distribution is \textit{narrower} than the bulk.
    	\textbf{Bottom Right:} Probability distribution of the ellipticity $\varepsilon_2$ in the $20-30\%$ bin, scaled by the mean to illustrate the width.  Note that the strangeness distribution is \textit{broader} than the bulk. Taken from \cite{Carzon:2019qja}.
    	}
    	\label{f:IBSQ42Centrality100K}
    \end{figure}
    %

It is clear from Figs. \ref{f:IBSQCentrality}-\ref{f:IBSQdistro} that strange quarks are especially sensitive to fluctuations (and likely fluctuate more than light quarks).  In addition, we would anticipate that light quarks have nearly the same fluctuations as the full energy density.  To check this systematically, we study the ratio $\varepsilon_n \{4\} / \varepsilon_n \{2\}$ where as $\varepsilon_n \{4\} / \varepsilon_n \{2\}\rightarrow 1$ there are fewer fluctuations in the system. However, small values of $\varepsilon_n \{4\} / \varepsilon_n \{2\}$ indicate large fluctuations in the system.

Looking at the cumulant ratio $\varepsilon_n \{4\} / \varepsilon_n \{2\}$ for the various charge and energy distributions, shown in Fig.~\ref{f:IBSQ42Centrality100K}, can tell us about the variance of the $\varepsilon_n$ distribution which is measured by the deviation of the ratio from unity. We can analyze this differentially by plotting the $\varepsilon_n$ probability distribution normalized by the mean to better reflect differences in width of the distributions corresponding to the deviation of the cumulant ratio from unity. For $< 70\%$ centrality in the first panel of Fig.~\ref{f:IBSQ42Centrality100K}, we see the baryon/electric charge distributions track the energy density profile quite closely, while strangeness is significantly different. There is an interesting crossing of curves at around $10\%$ centrality, where the strangeness distribution changes from having a ratio smaller than the bulk, in mid-central collisions, to a value larger than the bulk, in central collisions. We can investigate this more by plotting the distributions normalized by the mean for the $0-10\%$ centrality bin (pink band, second panel of Fig.~\ref{f:IBSQ42Centrality100K}) and the $20-30\%$ range (blue band, last panel of Fig.~\ref{f:IBSQ42Centrality100K}). For $20-30\%$, we see that the strangeness distribution is broader than the bulk, which reflects the increased number of event-by-event fluctuations due to a small number of strange quarks being produced. Event-by-event fluctuations of energy density are comparatively narrower and more peaked around their mean value, dictated by the strong mean-field background of the bulk elliptical geometry. In $0-10\%$ centrality, the strange quark distribution is narrower than the bulk, which reflects a qualitative change of the strangeness distribution relative to the bulk. Looking back at the $\varepsilon_2$ cumulant ratio, fluctuations of the energy density increase greatly when going to central collisions, which may be understood as the disappearance of mean-field elliptical geometry when the impact parameter goes toward zero. Event-by-event fluctuations, compared against the vanishing background, are greatly magnified which results in a broadening of the width of the $\varepsilon_2$ histogram. However, the strangeness distribution is broadened significantly less by this effect, resulting in a reduced slope for the strange $\varepsilon_2$ cumulant ratio and a crossing of the curves at $10\%$ centrality.  

The qualitative change in the strangeness distribution relative to the energy distribution as a function of centrality indicates, that in addition to a dependence on the number of quark pairs produced, the underlying geometries from which quarks are produced are different and act differently with respect to centrality. If the only dependence of the BSQ charge geometries was the number of particles produced, then we would expect to see the same hierarchy in $\varepsilon_2 \{4\} / \varepsilon_2 \{2\}$, that is, $E \approx B^+ \approx Q^+ > S^+$ maintained across mid-central and central collisions. 

The fact that the disappearance of the bulk elliptical geometry in central collisions affects the strangeness distribution differently, comes from fact that the geometry can produce strange quarks pairs that are not identical to the bulk geometry.  It is reasonable to assume that since the strange quarks have a large mass threshold of $2 m_s \approx 200 \, MeV$, they may only be produced by hot spots in the collision that contain enough energy density to meet this mass threshold. This can explain why the cumulant ratio of strangeness responds in a less singular manner than the bulk in central collisions: while the geometry of the bulk is becoming very round in central collisions, the geometry of the hot spots capable of producing strange quarks pairs is innately lumpier. Thus, the width of the strangeness distribution is less sensitive to the absence and the presents of a large elliptical bulk geometry.  

    %
    \begin{figure}
    \begin{centering}
      \includegraphics[width=0.45 \textwidth]{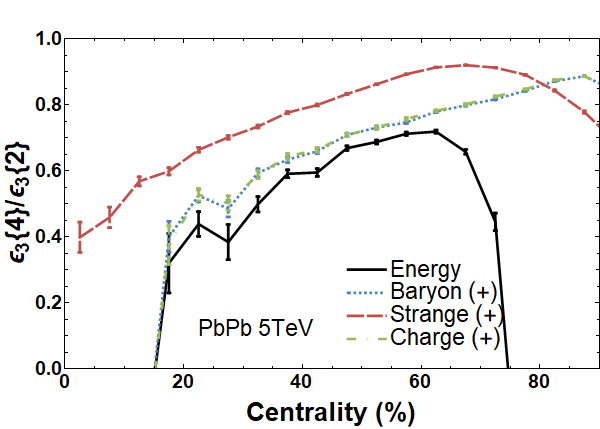} \,  \includegraphics[width=0.45 \textwidth]{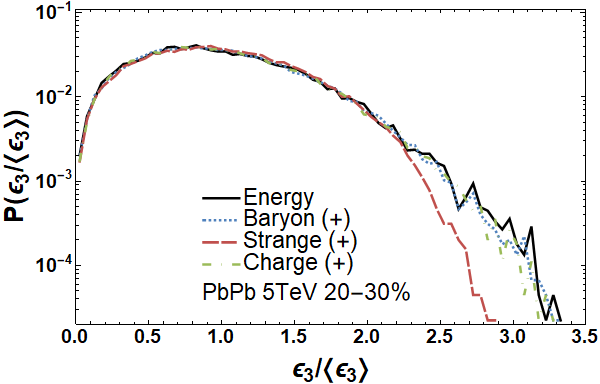}
   \end{centering}
    	\caption{
    	 Cumulant ratio $\varepsilon_3 \{4\} / \varepsilon_3 \{2\}$ versus centrality (left) and the probability distribution for $\varepsilon_3$ for the centrality class $20-30\%$ (right). Taken from \cite{Carzon:2019qja}.
    	}
    	\label{f:E3_IBSQ42Centrality100K}
    \end{figure}
    %

Looking at the cumulant ratio for triangularity, $\varepsilon_3 \{4\} / \varepsilon_3 \{2\}$ shown in the first panel of Fig.~\ref{f:E3_IBSQ42Centrality100K}, the absence of a mean-field background geometry produces an ordering hierarchy comparable to the one seen in the $0-10\%$ bin for $\varepsilon_2$: $S^+ > B^+ \approx Q^+ \approx E$. In central collisions, the strong elliptical shape of the event geometry, $\varepsilon_2$, disappears and the event-by-event fluctuations produce a large fractional change in the ellipticity, whose expectation value is close to zero. In a similar way, $\varepsilon_3$ has an expectation value close to zero, leading to large fractional changes compared to that small value. This is true for the gluons and light quarks which saturate the bulk geometry, but this is not the case for the strange quarks. Since fewer strange quarks are produced, they have an inherently lumpier hot spot distribution whose triangularity does not average automatically to zero. Thus both for $\varepsilon_2$ in the central bin of Fig.~\ref{f:IBSQ42Centrality100K} and for $\varepsilon_3$ across all centralities, the relative fluctuations of strange particles are smaller than the light quarks. We can further illustrate this through the probability distribution of $\varepsilon_3$ in the centrality window of $20-30\%$ in the second panel of Fig.~\ref{f:E3_IBSQ42Centrality100K} wherein strangeness, indeed, has a narrower distribution.  We argue that, in full context, these descriptors of the event-by-event fluctuations of the strangeness distribution and its centrality dependence provide systematic evidence for the coupling of strangeness to an independent underlying event geometry.

%
\section{Conclusion and Outlook} \label{Sec:Conclusion}
%

The ICCING algorithm provides a modular framework in which to introduce conserved charges to the initial state of heavy ion collisions through a Monte Carlo sampling of $g \rightarrow q\bar{q}$ splitting probabilities. ICCING was built with a model agnostic approach in mind and so has support for different input/output variables and formats as well as identification of areas where different model assumptions could be made. We have shown that the modification to the initial energy density, taken as input, is minimal and should not effect previous agreement with experimental results. Using eccentricity cummulants, it is possible to see that the strange charge follows a different geometry than the baryon/charge, which largely follows the energy geometry except at extremes. Investigation of cumulant ratios provides further evidence for the difference in behaviour between strange and baryon/charge densities.

\section*{Acknowledgements}
P.C. and J.N.H acknowledge support from the US-DOE Nuclear Science Grant No. DE-SC0020633, and from the Illinois Campus Cluster, a computing resource that is operated by the Illinois Campus Cluster Program (ICCP) in conjunction with the National Center for Supercomputing Applications (NCSA), and which is supported by funds from the University of Illinois at Urbana-Champaign. M. M. was supported in part by the US Department of Energy Grant No. DE-FG02-03ER41260 and BEST (Beam Energy Scan Theory) DOE Topical Collaboration.  M.D.S. acknowledges support from a startup grant from New Mexico State University.  D.E.W. acknowledges previous support from the Zuckerman STEM Leadership Program.

\bibliography{bibliography}

\end{document}